\documentclass[]{article}
\usepackage{graphicx}                    % For eps figures, newer & more powerfull
\usepackage[square]{natbib}
\begin{document}
\title{Millisecond solar radio bursts in the metric wavelength range}

\author{J. Magdaleni\'c\\ \emph{Hvar Observatory, Faculty of Geodesy, University of Zagreb}\\
A. Hillaris\\ \emph{University of Athens, Greece}\\
P. Zlobec\\ \emph{INAF-Trieste Astronomical Observatory, Italy}\\
B. Vr\v snak\\ \emph{Hvar Observatory, Faculty of Geodesy, University of Zagreb}}
\maketitle
\begin{abstract}
A study and classification of super-short structures (SSSs) recorded during metric type IV bursts is presented.
The most important property of SSSs is their duration, at half power ranging from 4-50 ms, what is up to 10 times
shorter than spikes at corresponding frequencies. The solar origin of the SSSs is confirmed by one-to-one
correspondence between spectral recordings of Artemis-IV\footnote{Thermopylae--Greece} 
and high time resolution single frequency measurements of the TSRS\footnote{Trieste--Italy}.

We have divided the SSSs in the following categories:
\begin{enumerate}
\item{Broad-Band SSSs: They were partitioned in two subcategories, the
\emph{SSS-Pulses} and \emph{Drifting SSSs};}
\item{Narrow-band: They appear either as \emph{Spike-Like SSSs} or as \emph{Patch-Like
SSSs};}
\item{Complex SSS: They consist of the absorption-emission segments and were morphologically subdivided
into \emph{Rain-drop Bursts} (narrow-band emission \emph{head} and
a broad-band absorption \emph{tail}) and \emph{Blinkers}.}
\end{enumerate}
\end{abstract}

\section{The \emph{Shortest}, as yet, observed Solar Radio Emissions}
%-------------------------------------------------------------------------------------
\begin{figure}[h!]
\centerline{\includegraphics[width=\textwidth]{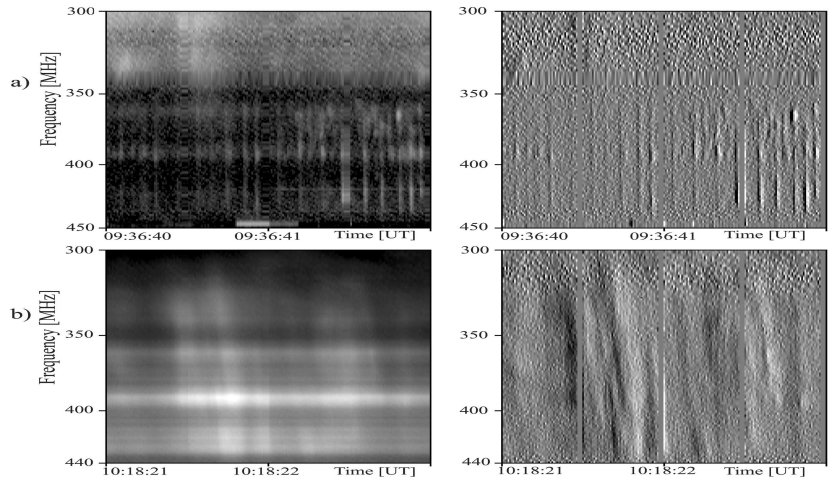}}
  \caption{ARTEMIS-IV Dynamic Spectra (Intensity on the left and differential on the right)
    of broad-band SSS: (a) SSS pulses, (b) Drifting SSSs.}
\label{Bsss}
\end{figure}
%-------------------------------------------------------------------------------------
\begin{figure}[h!]
\centerline{\includegraphics[width=\textwidth]{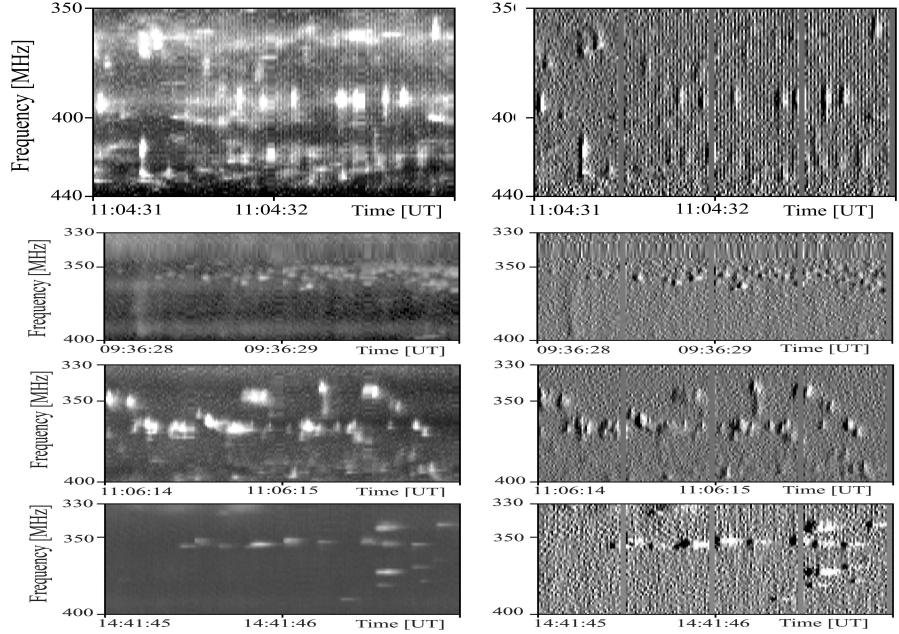}}
  \caption{ARTEMIS-IV Dynamic Spectra of narrow band SSS. Top to Bottom: 
Spike--like SSSs, dot--like, sail--like, flag--like. Intensity spectra are on the left 
and differential on the right.}
\label{Narrow}
\end{figure}
%-------------------------------------------------------------------------------------
\begin{figure}[h!]
\centerline{\includegraphics[width=\textwidth]{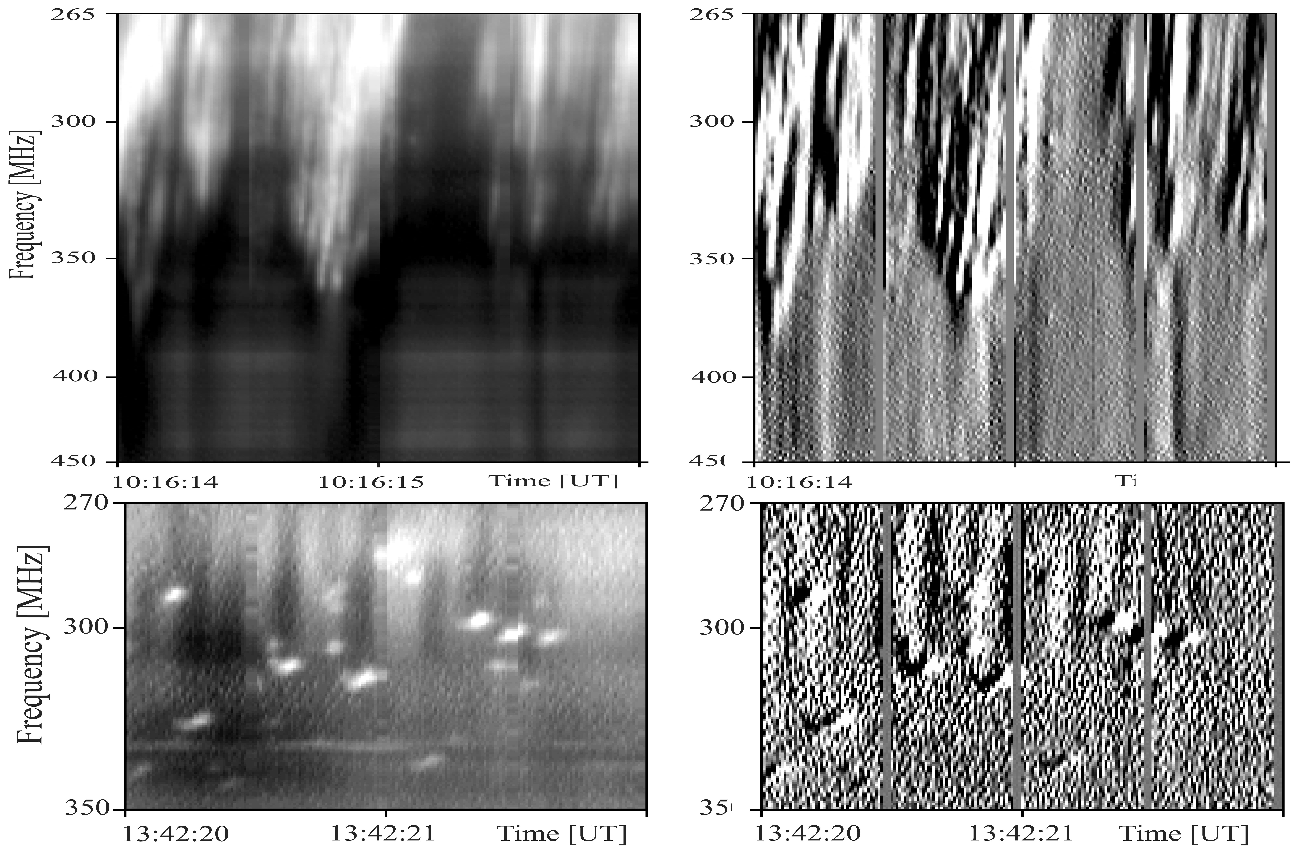}}
  \caption{ARTEMIS-IV Dynamic Spectrum of Blinker (Upper Panel) and Raindrop-like SSS (Lower Panel).
   Intensity spectra are on the left and differential on the right.}
\label{Complex}
\end{figure}
%-------------------------------------------------------------------------------------
The decimetric spikes (cf. \cite{Benz} for a review, also \cite{Fleishman_Melnikov}) have been long considered as the 
solar radio bursts with the shortest duration recorded; this has only
been occasionally challenged (\cite{Droege}, \cite{McConnell},
\cite{Elgaroy}). We present observational evidence and characteristics of a variety of solar radio bursts with
durations shorter than spikes up to an order of magnitude. These are, hence, named \emph{Super Short Structures}, or 
\emph{SSS}. Our observations consist of radio data with 1-10 ms time resolution, which enable the detection of, as yet 
unobserved, fine structure embedded in the type IV continuum; they comprise in particular :

\begin{itemize}
\item{Single frequency measurements at 237, 327, 408 610, 1420
and 2695 MHz, recorded with time resolution of 1 ms by the solar
multichannel radiopolarimetric system of the INAF-Trieste
Astronomical Observatory (TSRS-Trieste Solar Radio System).}

\item{ARTEMIS-IV, cf. \cite{Caroubalos}, dynamic spectra were obtained by the high sensitivity multichannel acoustooptical analyser (SAO) which covers the 265-450 MHz range, with time resolution of 10 ms.}

\end{itemize}

%-------------------------------------------------------------------------------------
\begin{table}[h!]
\begin{tabular}{{llll}} %
\hline\hline
\textbf{Class}  &\textbf{Subclass}& \textbf{Characteristics}                	&\textbf{Morphologically} \\
                &           &                               			&\textbf{Similar Bursts} \\
\hline\hline
 Simple Broad   &   SSS     & $\Delta f \approx$ 100MHz,~Duration=10-20 ms    & Pulsations \\
 Band SSS       &   Pulses  & Some Groups Exhibit  Quasi Periodicity		&        \\
                &           &                                   			&        \\
                &  Drifting & $|\Delta f / \Delta t| \approx$ 400-1000 MHz/sec & Type III like Bursts \cite{Isliker}\\
                &  SSS  & Duration=30-70 ms, $\Delta$ f $\ge$ 100MHz        	& \\
\hline\hline
Simple Narrow   & Spike-like& $\Delta f  \le$   20 MHz, $|\Delta f / \Delta t| \ge$ 800 MHz/sec & Spikes \cite{Benz}\\
Band SSS        &   SSS 	& Duration=4-30 ms                      		&  \\
                &           	&                                   		&        \\
                & Patch-like	& $\Delta f  \le$   15 MHz,~Duration=4-50 ms    &       \\
                &  SSS  	& Further Subdivision:                      	&        \\
                &       	& Dot-like SSS                  			& Dot like Structures \cite{Sawant} \\
                &           	& Sail-like SSS and Flag-like SSS   		& -      \\
\hline\hline
Complex SSS     &  Blinkers 	& $|\Delta f / \Delta t| \approx$ 650 MHz/sec   & -  \\
 with emission  &       	& $\Delta f~>$ 150 MHz,~Duration=30-40 ms     	&        \\
 and absorbtion &           	& Absorption switches to emission               &        \\
 element        &           	&                                   		&        \\
                & Raindrop  	& Emission \emph{HEAD},~Duration$\approx$50 ms  & Tadpole  Bursts \cite{Slottje} \\
                & bursts    	& $\Delta f \approx 5 MHz$, $|\Delta  f / \Delta  t| \approx 60 \pm10$ MHz/sec &     \\
                &           	& Absorption \emph{TAIL},~Duration$\approx$30 ms    		&        \\
                &           	& $\Delta f \approx 40 MHz$, 				&        \\
                &       	&  $|\Delta  f/\Delta t|\approx 1000 \pm 400$ MHz/sec &        \\
\hline\hline
\end{tabular}
\caption{Classification and Basic Characteristics of Super Short Structures (\textbf{SSS})}
\label{tab}
\end{table}
%-------------------------------------------------------------------------------------
The credibility of the \emph{SSSs} was confirmed by one-to-one
identification of individual SSS bursts in the single frequency
recordings and in the corresponding ARTEMIS-IV spectra. 
We have classified the \emph{SSSs} in the following morphological categories:
\begin{itemize}
\item{\emph{Simple broad-band SSSs}; characterized by a broad frequency bandwidth
	$\Delta f \geq$ 100 MHz (Figure \ref{Bsss}). They can be subdivided in two subcategories:}
	\begin{itemize}
		\item { \emph{SSS-pulses} have duration in the range
		10-20 ms, and frequency bandwidth $\approx$100 MHz. They appear in groups and, occasionally,
		exhibit quasi-periodic behaviour. Due to morphological similarity with 
		pulsations of the type IV fine structure, they may be considered as an extension of 
		pulsating structures towards higher time scales.}
		\item  { \emph{Drifting SSSs} have duration in the range
		30-70 ms. Their bandwidth in general exceeds the 100 MHz, and frequency 
		drift rates  ($|\Delta f / \Delta t| \approx$ 400-1000 MHz/sec)
		are similar to the drift rates of the metric type III bursts. (\cite{Isliker}, \cite{Gudel}).}
	\end{itemize}
\item{\emph{Simple narrow-band SSSs} are distinguished by their narrow frequency bandwidth
$\Delta f \le$ 20MHz. They are, also, subdivided into two subcategories:}
	\begin{itemize}
		\item {\emph{Spike-like SSSs} are  the shortest \emph{SSSs} with a duration in the 4-30 ms range 
		(Figure \ref{Narrow} top panel). Their frequency bandwidth is mostly $\Delta f  \le$ 20 MHz.
		If measurable, frequency drifts are $|\Delta f / \Delta t| \ge$	800 MHz/sec. 
		On dynamic spectra they very much resemble spikes,~(\cite{Benz}) as their name implies.}

		\item {\emph{Patch--like SSSs} exhibit, due to their morphological diversity,
		a rather broad range of duration which varies between 4 and 50 ms (Figure \ref{Narrow}, 
		three lower panels). 
		The frequency  bandwidth is $\Delta f  \le$ 15 MHz, and it can be as low as a few
		MHz. This qualifies \emph{patch--like SSS} as the \emph{SSSs}
		of the narrowest bandwidth. Their spectral appearance varies; 
		they can resemble dots, sails or flags and were
		further subdivided accordingly to:  \emph{dot-like SSS},
		\emph{sail-like SSS} and \emph{flag-like SSS}. \emph{Dot-Like} 
		structures in the 1000-2000 MHz range, recorded by
		the Brazilian Solar Spectroscope with 50 ms resolution, have been reported by \cite{Sawant}.}

	\end{itemize}
\item{\emph{Complex Super Short Structures} are characterised by an emission and an absorption
element. Two subcategories could be distinguished:}
	\begin{itemize}
		\item {\emph{Rain-drop bursts} (Figure \ref{Complex}, lower panel)
		consist of a narrow-band emission \emph{HEAD} ($\Delta f \approx 5 MHz$)
		and a broad-band absorption \emph{TAIL} ($\Delta f \approx 40 MHz$). 
		The durations are approximately 50 ms for
		the \emph{HEAD} and 30 ms for the \emph{TAIL}. Both the \emph{HEAD} and the
		\emph{TAIL} exhibit frequency drift which is 
		$|\Delta  f / \Delta t| \approx 60 \pm10$ MHz/sec 
		and $|\Delta  f/\Delta t|\approx 1000 \pm 400$ MHz/sec, respectively. 
		Morphologically these \emph{SSSs}
		resemble, somehow, to the tadpole bursts (\cite{Slottje}).}

		\item {\emph{Blinkers} (Figure \ref{Complex}, upper panel)
		are drifting bursts ($|\Delta f / \Delta t| \approx$ 650 MHz/sec)
		consisting of absorption element that is switching abruptly to
		emission element. Opposite cases have also been found (emission in
		the high-frequency part, and absorption in low-frequency part of
		the burst). The duration $d_{1/2}$=30--40 ms, is approximately the
		same along the whole burst. They are the \emph{SSSs} with the largest
		frequency bandwidth $\Delta f ~>$ 150 MHz. }
	\end{itemize}
\end{itemize}
\section{Discussion \& Conclusions}
The analysis of high time resolution spectral recordings
(ARTEMIS-IV) and single frequency measurements (TSRS-data) reveals
a number of different classes of \emph{Super Short Structures}.
The basic characteristics of the described \emph{SSS} classes are
summarized in Table \ref{tab}.

It is stressed that all of the presented features have
duration considerably shorter than spikes which in the frequency
range 250-450~MHz have duration to 100--50 ms. This does
not necessarily establishes \emph{SSSs} as the shortest in
existence, but only as the shortest recorded until now. It is to be expected
that the improvement of instrumental
resolution may reveal the existence of even shorter radio bursts.

The duration of the \emph{SSSs} varies with frequency, and from event to event.
Therefore, at present  it is not possible to establish a systematical dependence of the
\emph{SSSs} duration on the observation frequency; this would be in
favour of plasma radiation mechanisms which cannot be yet confirmed.
For more specific results a detailed case study is under way.

Lastly, although some morphological similarities with already
known burst types (pulsations, type IIIs, tadpoles) exist, there
is no as yet conclusive evidence of a common radiation mechanism.

\end{document}